\documentclass[runningheads]{llncs}
\usepackage[T1]{fontenc}
\usepackage{subfiles}
\usepackage{tabularx}
\usepackage{hyperref}
\usepackage{graphicx}
\usepackage{xcolor}
\usepackage{array}
\usepackage{amsmath,amssymb,amsfonts}%
\usepackage{mathrsfs}%
\newcolumntype{C}{>{\centering\arraybackslash}X}
\newcolumntype{P}[1]{>{\centering\arraybackslash}p{#1}}

\usepackage{listings}
\usepackage{lipsum}
\usepackage{xcolor}
\usepackage{eqparbox}
\usepackage{verbatimbox}

\definecolor{codegray}{gray}{0.9}

\lstdefinestyle{groovytest}{
  backgroundcolor=\color{codegray},        
  basicstyle=\ttfamily\small,
  keywordstyle=\color{blue}\bfseries,
  commentstyle=\color{gray}\itshape,
  stringstyle=\color{orange},
  showstringspaces=false,
  columns=fullflexible,
  keepspaces=true,
  frame=none,                              
  captionpos=b,
  numbers=left,                            
  numberstyle=\tiny\color{gray},           
  numbersep=10pt                           
}

\begin{document}
\title{On the Marriage of Theory and Practice in Data-Aware Business Processes via Low-Code - Extended version}

\author{Ali Nour Eldin\inst{1,2}, Benjamin Dalmas\inst{2}, Walid Gaaloul\inst{1}}

\authorrunning{A. Nour Eldin}

\institute{Telecom SudParis, Institut Polytechnique de Paris, France, \\ email: \email{\{firstname.last\_name\}@telecom-sudparis.eu}  \and Bonitasoft, France, email: \email{\{firstname.last-name\}@bonitasoft.com}
}

\maketitle           

\begin{abstract}
In recent years, there has been a growing interest in the verification of business process models. Despite their lack of formal characterization, these models are widely adopted in both industry and academia. To address this issue, formalizing the execution semantics of business process modeling languages is essential. Since data and process are two facets of the same coin, and data are critical elements in the execution of process models, this work introduces Proving an eXecutable BPMN injected with data, BPMN-ProX. BPMN-ProX is a low-code testing framework that significantly enhances the verification of data-aware BPMN. This low-code platform helps bridge the gap between non-technical experts and professionals by proposing a tool that integrates advanced data handling and employs a robust verification mechanism through state-of-the-art model checkers. This innovative approach combines theoretical verification with practical modeling, fostering more agile, reliable, and user-centric business process management.

\end{abstract}

\section{Introduction}

With the substantial growth in the size and complexity of business processes due to technological advancements and increased regulatory requirements~\cite{DBLP:conf/ant/PaivaFFM18}, ensuring the correctness and compliance of these processes with various requirements has become a significant challenge~\cite{DBLP:journals/corr/BohmerR15}. Furthermore, in an environment where continuous adaptation is crucial, the importance of testing business processes to ensure their continued value and adaptability is increasingly critical~\cite{DBLP:journals/bpmj/Guerreiro21}.

A business process in an organization comprises interconnected activities arranged to achieve a specific goal. The process and data views are essential for understanding these processes. The process perspective, or \textit{control-flow perspective}, outlines task sequences using BPMN~\cite{bpmn_specification}. The data perspective, similar to UML~\cite{omg2011umls}, focuses on business and processing data. Integrating these views into a unified workflow modeling language enhances information representation, reduces implementation time, and supports decision-making and resource allocation.

Organizations consistently face challenges in integrating data and processes~\cite{DBLP:conf/otm/Hull08,DBLP:conf/otm/Reichert12,DBLP:SLR}, which include multiple complexities. Firstly, the modeling challenge involves developing a robust framework to accurately represent intricate interactions and data manipulations within business processes. Secondly, verification ensures that these models comply with operational standards, some works proposed a verification frameworks for the process enriched by data. Thirdly, execution pertains to the practical application and monitoring of these models, ensuring they function effectively in real-time operational environments. 
An additional challenge is the selection of modeling constructs, which often differ from those provided by data and process modeling standards such as BPMN, SQL, and UML. 
Lastly, the automation and verification of such integrated models are complex for designers, often requiring assistance from developers or experts to successfully implement, verify, and automate a business model due to the lack of comprehensive tools that can address all aspects of the integration process.

These dimensions can be recognized in their full complexity when it comes to the \textit{verification} of the resulting integrated models~\cite{DBLP:conf/pods/CalvaneseGM13,DBLP:journals/siglog/DeutschHLV18}. Verification is crucial because data and process models that seem correct in isolation may produce errors once integrated\cite{DBLP:conf/caise/PolyvyanyyWOB19}. The verification involves converting the utilized language into a well-established standard for data and process components. Specifically, this means that either the control flow is represented using Petri nets or other mathematical formalisms suitable for dynamic systems that are not easily comprehensible through front-end notations like BPMN, or the data manipulation aspects depend on abstract, logical operations that are not directly expressible in concrete data manipulation languages like SQL or OCL. Verification is crucial as it ensures the rigor and theoretical validity of executable models, essential for their flawless practical application. This step not only confirms that the models adhere to operational standards but also prevents costly errors and malfunctions before implementation. A last crucial issue is that the vast majority of the literature in this spectrum mainly provides foundational results that do not directly translate into effective verification tools. Therefore, it is crucial to have a tool that assists designers and non-technical experts in verifying the model before automating its execution. Such a tool can help detect errors and reduce maintenance time prior to the execution phase. Supporting these dimensions covers the first three parts of the business process life cycle —design, implementation, and execution— by ensuring the integration of data and processes into a verified executable model.

This work addresses these limitations by introducing BPMN-ProX, a Low-Code Testing Framework designed to prove and verify executable BPMN models enriched with data. Since the goal of the low-code solution is to minimize manual coding and enable non-technical experts to create applications easily, this framework combines data and process modeling with theoretical verification and practical execution, ensuring the integrity and compliance of models. This approach allows non-technical experts to verify and execute BPMN with minimal assistance from developers. The framework is based on BPMN and UML standards for data modeling, with the artifacts of these languages translated into SQL standards to facilitate integration with verification techniques.

BPMN-ProX extends an existing low-code framework to support graphical data representation within BPMN, which serves as the front-end modeling language based on the definitions from previous studies~\cite{NourEldin2024LowCode}. This paper presents how this language instantiates the data-related aspects of the abstract modeling language studied in~\cite{data-aware-BPMN}. The features of BPMN-ProX are derived from the requirements for concrete, executable, and verifiable data-aware process modeling languages distilled from the literature.

Furthermore, an implementation of BPMN-ProX in real business process management systems is demonstrated, showing how existing artifacts can be converted to an SQL-based language. Additionally, a translator that converts a BPMN-ProX model into the syntax of MCMT~\footnote{http://users.mat.unimi.it/users/ghilardi/mcmt/} based on the encoding rules defined in previous studies~\cite{data-aware-BPMN} is reported. MCMT is a state-of-the-art SMT-based model checker for infinite-state systems that can be used for verification. 

The remainder of this paper is organized as follows: Section~\ref{sec:related_works} explores the existing literature. Section~\ref{sec:preliminaries} presents the preliminaries upon which our paper is based. In Section~\ref{sec:BPMN_extension}, we introduce our framework, BPMN-ProX. This is followed by a discussion of the the tools used for verification our model in Section~\ref{sec:verification_techniques}. Finally, Section~\ref{sec:conclusion} concludes the paper.

\section{Requirement Analysis and Related Work}
\label{sec:related_works}
\label{sec:vdfbpmn:related_work}
The integration of data and processes is a longstanding area of research at the intersection of BPM, data management, process mining, and formal methods. Given the focus on verification and the integration of verification methods within an executable process, the discussion is confined to relevant works addressing process and data modeling, process and data verification, and process and data execution in BPMN.

Table~\ref{tab:related_work} outlines six key requirements (R1–R6) for modeling languages used in the verification, execution, and deployment of executable models within Business Process Management Systems (BPMSs). Requirements R1, R5, and R6 are inspired by prior works focused on the integration of process and data modeling and execution~\cite{dfbpmn,NourEldin2024LowCode,DBLP:delta_BPMN}, while requirements R2 to R4 are derived from foundational studies on verification frameworks~\cite{DBLP:delta_BPMN,DBLP:journals/pvldb/LiDV17}. These six requirements capture critical capabilities expected from modeling approaches, including support for expressive data modeling, automated verification, property analysis, data handling, model executability, and integration within low-code or industrial platforms. Table~\ref{tab:related_work} compares the extent to which various approaches meet these requirements, using the notation (+) for full support, (+/-) for partial support, and (–) for lack of support.

\begin{table}[h]
\centering
    \begin{tabularx}{1.05\textwidth}
    {|P{5cm}|C|C|C|C|C|C|c|} 
    \hline
Framework & R1  & R2  & R3  & R4 & R5 & R6  & Verification Framework  \\ \hline
Acticity View\cite{DBLP:activity-view}, DF-BPMN \cite{dfbpmn}                                                                  & +/- & -   & -   & -  & -  & -   & -                       \\ \hline
Linking Data and BPMN\cite{DBLP:conf/caise/GiacomoOET17}, Complex Data Dependencies\cite{DBLP:conf/bpm/MeyerPFW13} & +/- & -   & -   & -  & +  & -   & -                       \\ \hline
Symbolic BPMN verification\cite{DBLP:conf/wrla/DuranRS18}                                                                                 & +/- & -   & +/- & +  & -  & -   & symbolic verification   \\ \hline
data-aware BPMN\cite{data-aware-BPMN}                                                                                                     & +/- & +   & +   & +  & -  & -   & data-aware safety       \\ \hline
VERIFAS\cite{DBLP:journals/pvldb/LiDV17}                                                                                                  & +/- & +   & +   & +  & -  & -   & fragment of LTL-FO      \\ \hline
BAUML\cite{DBLP:BAUML}                                                                                                                    & +/- & +   & +/- & -  & -  & -   & fixed test cases        \\ \hline
ISML\cite{DBLP:conf/caise/PolyvyanyyWOB19}                                                                                                & +/- & +/- & -   & -  & +/-  & -   & state-space exploration \\ \hline
dapSL\cite{DBLP:conf/caise/CalvaneseMPR19}                                                                                                & +/- & +/- & -   & -  & +/-  & +   & state-space exploration \\ \hline
delta-BPMN (PDMML) \cite{DBLP:delta_BPMN}                                                                                                      & +/- & +   & +   & +  & -  & +/- & data-aware safety       \\ \hline
BPDML\cite{NourEldin2024LowCode}                                                                                                                & +   & -   & -   & -  & +  & +   & -                       \\ \hline
\textbf{BPMN-ProX (ours)} & +   & +   & +   & +  & +  & +   & data-aware safety       \\ \hline
\end{tabularx}
\caption{Requirements coverage (covered (+), partially (+/-), lack of support (-))}
\label{tab:related_work}
\end{table}

\textbf{R1. Data modeling.} Data modeling is a crucial aspect of process modeling that encompasses various significant elements drawn from prior studies. In this paper, data modeling includes data injected by the external environment, interactions between data, and the execution of complex data operations. Within data-aware processes, it is essential to acknowledge that new data may be introduced during process execution. A critical consideration is how data access is managed, as this can influence the type of verification applied. Moreover, the interaction between data entities and the expressiveness of data operations enhances the clarity and comprehensibility of the models. These elements are integrated into our framework to address the multifaceted requirements of data-aware process modeling and verification, thus motivating their selection and inclusion in this work.

\textbf{R2. Tool support for verification.} Effective verification requires robust tool support to automate and simplify the process. Tools must offer functionalities such as model checking, state-space exploration, and satisfiability checking, enabling users to verify complex properties efficiently. Robust tool support automates verification tasks, reducing human error and enhancing efficiency. It enables scalable analysis of complex systems, identifies issues early, and supports continuous development processes, ensuring reliable and high-quality outcomes.

\textbf{R3. Key property analysis.} Verification approaches should rigorously analyze key properties, including soundness, completeness, and termination. Analyzing properties like soundness, completeness, and termination ensures process reliability and prevents errors such as infinite loops or incomplete executions. This rigorous validation is vital for system stability and user trust.

\textbf{R4. Separation of read-only and mutable data.} The language separates read-only persistent data from persistent data that are
updated during the execution. Separating read-only data from mutable data enhances security, optimizes performance, and simplifies data management. It reduces concurrency issues and supports efficient caching, making systems more robust and scalable.

\textbf{R5. Executable model.} One of the important aspects of proposing a new modeling language is execution. The execution phase enables academia and industry to interact with the proposed modeling language. Each approach suggests a fundamental method for executing their process model, either by transformation to Petri nets or by integration with existing business process systems.

\textbf{R6. Integration within BPMSs.} Lastly, the integration of these approaches within existing BPMSs is essential for incorporating research into industry, thereby enhancing the significance of the research.

The comparison in Table~\ref{tab:related_work} summarizes the key features of the discussed verification approaches. 
Activity View~\cite{DBLP:activity-view} and DF-BPMN~\cite{dfbpmn} are approaches limited to process and data modeling (R1), without addressing the execution or verification phases. Even within data modeling, their capabilities are restricted, as they do not support complex data operations such as arithmetic or mathematical computations.

Linking Data and BPMN~\cite{DBLP:conf/caise/GiacomoOET17} and Complex Data Dependency~\cite{DBLP:conf/bpm/MeyerPFW13} are approaches that support process modeling and execution (R1, R5). However, they require expert knowledge to execute the processes and to transform the modeling language into an executable format. Moreover, the authors do not consider verification, which can lead to undetected errors during process execution.

VERIFAS~\cite{DBLP:journals/pvldb/LiDV17} includes an embedded, ad-hoc verification tool that supports model checking of properties expressed in a fragment of first-order Linear Temporal Logic (LTL). While it provides strong tool support for verification (R2) and supports separation of read-only and mutable data (R4), its key property analysis (R3) is limited to specific classes of data-aware processes, which can affect completeness and termination guarantees. Furthermore, it lacks full support for expressive data modeling (R1) and complex data flows, potentially limiting its applicability in comprehensive data-aware scenarios.

BAUML~\cite{DBLP:BAUML} uses first-order satisfiability checking over temporal flows with fixed test cases to express verification properties. It provides partial tool support (R2) and offers a semi-decidable approach to key property analysis (R3), with no guarantee of termination in general. BAUML also shows limitations in modeling external and volatile data (R1) and does not offer a robust execution framework or integration with low-code platforms (R5, R6), reducing its practical utility for non-technical users.

ISML~\cite{DBLP:conf/caise/PolyvyanyyWOB19} is based on Colored Petri nets and assumes bounded data domains, but lacks a dedicated verification language. As a result, it provides only limited tool support (R2) and weak analysis of key properties (R3). It also struggles with advanced data modeling (R1) and does not clearly distinguish between mutable and read-only data (R4). These constraints restrict its ability to handle complex, real-world data-aware workflows.

DapSL~\cite{DBLP:conf/caise/CalvaneseMPR19} exhibits similar limitations, relying on ad-hoc state-space constructions with no formal verification framework. It provides basic tool support (R2) and partial support for key property analysis (R3), but lacks flexibility in modeling complex data interactions (R1) and offers no integrated support for mutable data handling (R4) or full execution capabilities (R5).

delta-BPMN~\cite{DBLP:delta_BPMN} integrates verification of data-aware safety properties into the MCMT model checker via the DAB framework~\cite{data-aware-BPMN}. It excels in tool support (R2) and offers formal guarantees for key properties like soundness, completeness, and termination (R3). It also supports the separation of data access semantics (R4). However, its lack of support for complex data operations (R1) and limited integration with low-code or executable environments (R5, R6) reduce its flexibility in industry settings.

BPDML~\cite{NourEldin2024LowCode} is a low-code solution designed for non-expert users; however, it lacks support for verification (R2–R4) and primarily focuses on modeling and execution of the process and data model.

\medbreak
This paper focuses on combining practical and theoretical process models injected with data using low-code solutions. Therefore, the starting point is BPDML, which is a low-code solution for executing data and models in any BPMS. BPDML is extended as BPMN-ProX, with the verification aspect addressed through transformation to MCMT. The soundness, completeness, and termination of the algorithmic technique implemented in MCMT are extensively studied. Additionally, the complex code is verified by automating unit tests for the executable code.

\section{Background}
\label{sec:preliminaries}
Business process modeling is a foundational activity in digital transformation, enabling the design and automation of organizational workflows. While BPMN is widely adopted for modeling the control flow of processes, it offers limited support for representing the data manipulated throughout process execution~\cite{DBLP:SLR}. This gap creates practical challenges, as data is fundamental to most business decisions and interactions. In many real-world settings, analysts must supplement BPMN models with textual specifications or separate data models, leading to misinterpretations, increased collaboration overhead, and delays in process implementation~\cite{dfbpmn}.

Recognizing this fragmentation, recent research has highlighted the importance of integrating both data and process perspectives into a unified model~\cite{DBLP:SLR}. However, existing approaches are either too specialized (e.g., SQL-based internal models like dapSL~\cite{DBLP:conf/caise/CalvaneseMPR19}), too rigid (e.g., Petri net extensions~\cite{DBLP:dbnets}), or not designed for execution by non-technical users~\cite{DBLP:delta_BPMN}. Even commercial BPMS tools such as Bonita~\footnote{https://www.bonitasoft.com/} or Camunda~\footnote{https://camunda.com/} require developers to manually implement data logic using Groovy or Java, which introduces further complexity and technical barriers~\cite{DBLP:conf/bpm/MeyerPFW13}.

\begin{figure}[!ht]
\centering
\includegraphics[width=0.9\linewidth]{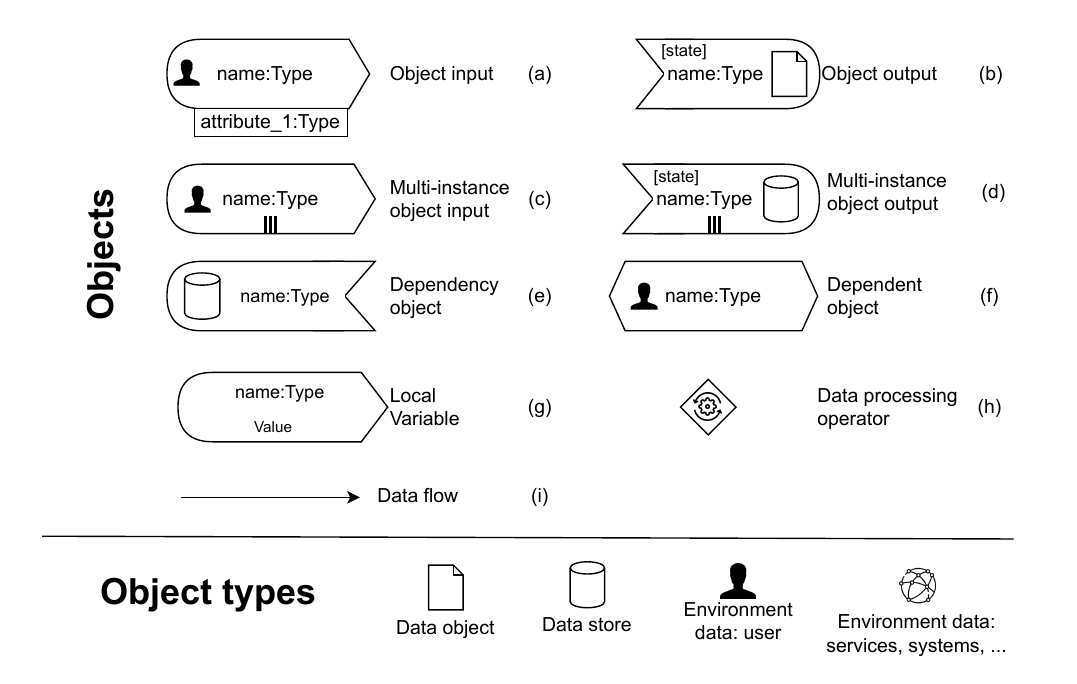}
\caption{Symbols used in BPDML.}
\label{fig:dfbpmn:dfbpmn:symbols}
\end{figure}

To address these limitations, BPDML (Business Process and Data Modeling Language) was introduced as a low-code extension of BPMN designed to bridge the gap between process logic and data semantics~\cite{dfbpmn,NourEldin2024LowCode}. This modeling language elevates data interactions to first-class citizens in process modeling, enabling users to define not only control flow but also the structure, flow, and operations on data within each activity. To achieve this, BPDML enriches BPMN activities with a zoomable sublayer that reveals detailed input and output data, their attributes, and the corresponding operations. These data elements are represented with intuitive graphical symbols—such as semi-circular connectors for inputs and outputs, icons for data stores, process variables, user forms, and external services—all annotated with data types, attribute names, and optional multiplicities, as illustrated in Figure~\ref{fig:dfbpmn:dfbpmn:symbols}.

\begin{figure}[!ht]
\includegraphics[width=\linewidth]{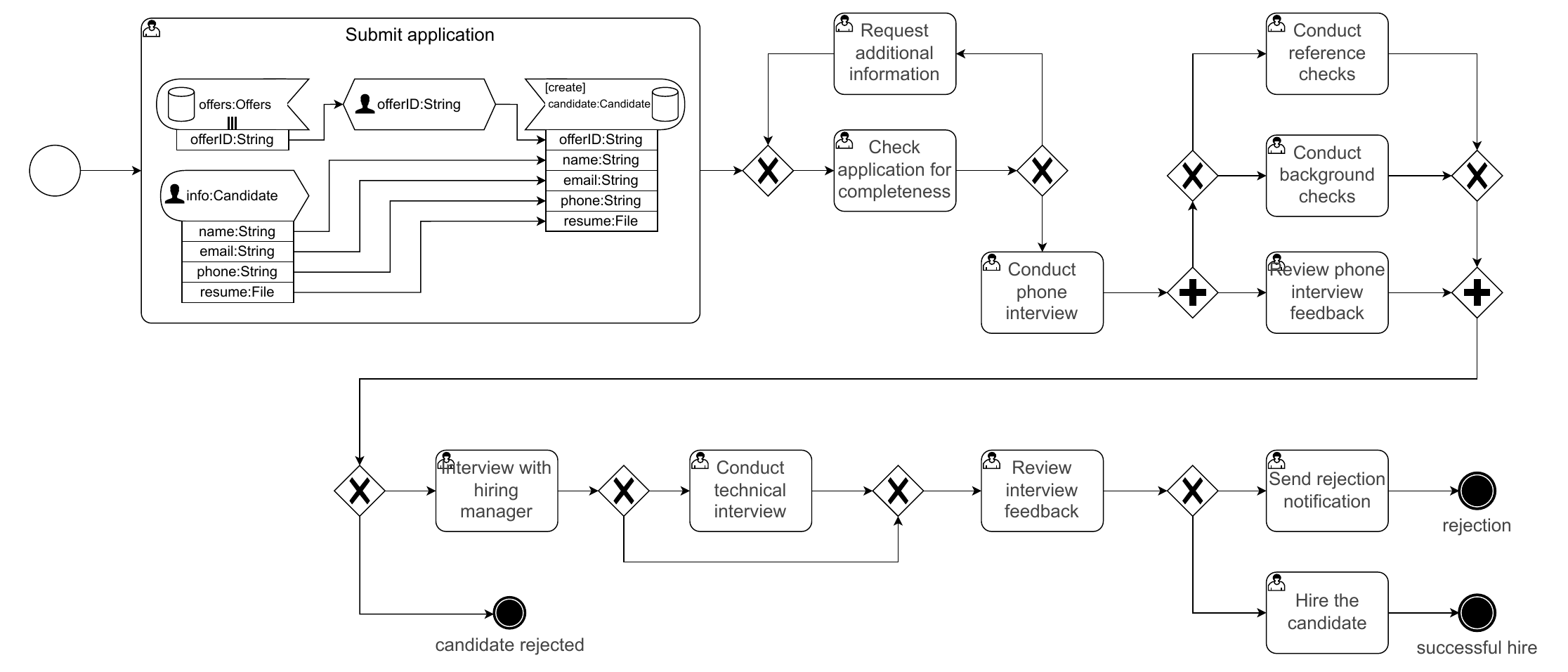}
\centering
\caption{Part of BPDML example which expand "Submit application" activity and collapse the others activities.}
\label{fig:dfbpmn:dfbpmn_full_example}
\end{figure}

Each output in BPDML is explicitly associated with a CRUD status such as \textit{[create]}, \textit{[update]}, \textit{[delete]}, or \textit{[read]}, providing clarity on how data is manipulated during execution. The relations between inputs, outputs, and intermediate logic are captured using data flow arrows, while more complex operations—like conditionals or aggregations—are abstracted using a ``data processing operator" symbol. For operations that cannot be easily expressed graphically, the model supports structured natural language descriptions using the Gherkin language. This structured format allows for clear, executable definitions of behaviors while avoiding the ambiguity of free-form text, making BPDML suitable for collaboration between business users, analysts, and developers. An example application of BPDML, showing an expanded ``Submit application" activity within a hiring process, is illustrated in Figure~\ref{fig:dfbpmn:dfbpmn_full_example}.

\section{BPMN-ProX: Proving an eXecutable BPMN injected by data}
\label{sec:BPMN_extension}

In this section, the proposed framework is defined. This is followed by the constraints established to control the modeling phase, and the encoding of the language to data-aware BPMN, which is then translated into SQL. Finally, the guard proposed for the conditional flows is defined.
\label{sec:vdfbpmn:verification}
This section presents our framework for verifying data and process aspects in BPMN-ProX. By aligning BPMN-ProX with the state-of-the-art data-aware verification approach~\cite{data-aware-BPMN}, we adopt a front-end modeling strategy that, in principle, mirrors Delta-BPMN’s idea of providing a modeling language~\cite{DBLP:delta_BPMN} to be verified against data-aware BPMN. However, whereas Delta-BPMN typically requires significant expertise, BPMN-ProX is designed with non-technical users in mind, offering more direct and data-rich modeling constructs that remain accessible to broader audiences. We begin by presenting an overview of our approach, which sets the foundation for verifying BPMN-ProX models.  Next, we identify the range of data sources supported in our extension and detail the constraints that guide the modeling phase. We then show how BPMN-ProX can be mapped to a data-aware BPMN specification—similar in spirit to what Delta-BPMN achieves—enabling a subsequent translation into SQL. We also describe the guard mechanism proposed for conditional flows. Lastly, we present the automation of unit tests.

\subsection{Approach overview}
\label{sec:vdfbpmn:overview}

\begin{figure}[!ht]
\centering
\includegraphics[width=1\linewidth]{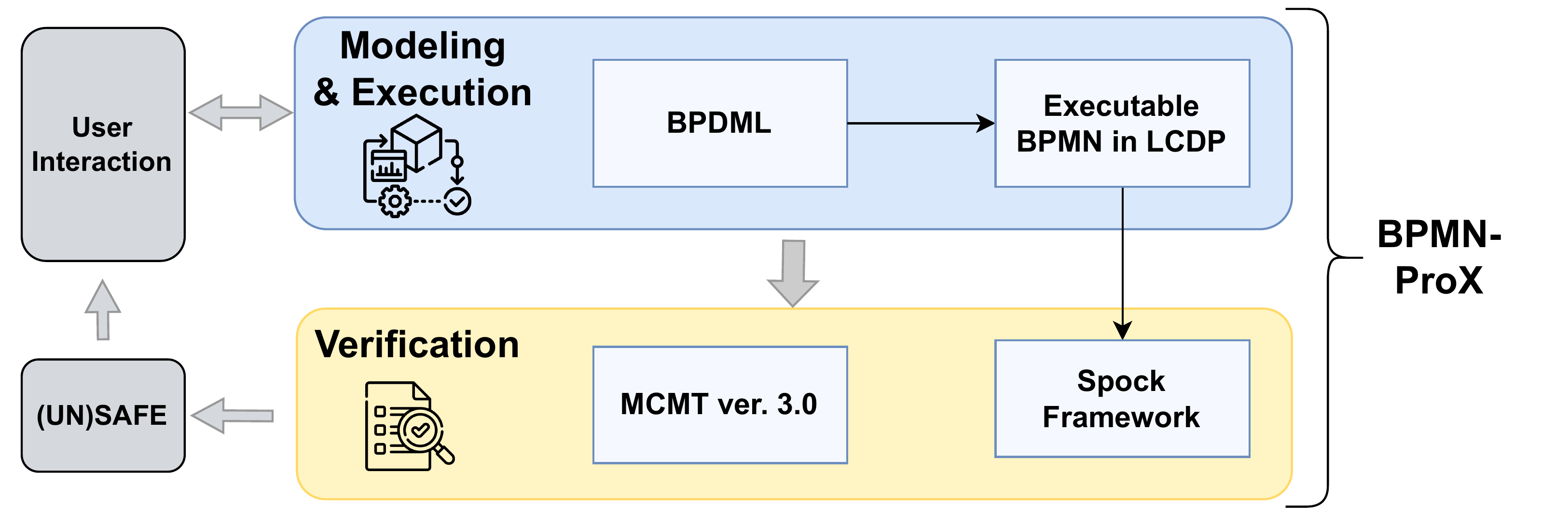}
\caption{General architecture of the proposed approach.}
\label{fig:vdfbpmn:overview}
\end{figure}

Figure~\ref{fig:vdfbpmn:overview} provides an overview of the proposed approach, illustrating the interaction between modeling, execution, and verification components, as well as the role of user interaction in ensuring process safety. The figure highlights the flow from BPDML modeling to executable scripts, along with the intended verification mechanisms using BPMN-ProX.

Our objective is to verify both the process and data within a model. Since our focus is on updatable data, the verification of data-based operations becomes a crucial aspect of our approach. Since updatable data is not just passive information—it directly influences decision points, control flow, and the state of business entities throughout the execution of a process. If operations on such data are incorrectly specified or executed, the consequences can propagate across the process lifecycle, leading to inconsistent behaviors, unexpected outcomes, or violations of business rules.

To address this, we transform BPDML models into MCMT specifications by leveraging the state-of-the-art techniques in data-aware BPMN modeling. This transformation allows us to formally reason about how processes interact with data, particularly when updates are involved.

However, given that MCMT does not support arithmetic expressions or complex operations, our framework goes beyond formal verification alone. To ensure a comprehensive verification strategy, we combine both formal and dynamic verification techniques. This is achieved through the integration of two complementary tools: MCMT and the Spock framework.

MCMT is particularly effective for the formal verification of infinite-state systems, especially in scenarios where correctness of data and control flow needs to be guaranteed through exhaustive logical reasoning~\cite{data-aware-BPMN}. Its compatibility with array theories and logical constraints makes it well-suited for verifying BPDML models where updatable data plays a central role.

In parallel, the Spock framework\footnote{\url{https://github.com/spockframework/spock}} is employed to handle runtime validation of complex and dynamic behaviors that are difficult to capture formally. As a behavior-driven testing framework for Groovy, Spock aligns seamlessly with our low-code execution environment (such as Bonita), enabling expressive unit testing of scripts and process logic. This dual-verification approach ensures both theoretical soundness and practical correctness throughout the process lifecycle.

The modeling component of our approach is built upon the BPDML framework introduced in Section~\ref{sec:preliminaries}, which we extend to support read-only data properties. These properties serve as pre-execution alerts to enhance data integrity, detailed in Section~\ref{sec:vdfbpmn:verification:data_source}. Additionally, syntactic validations embedded in BPMN-ProX facilitate early error detection and guide users in model construction, as elaborated in Section~\ref{sec:vdfbpmn:verification:syntax}.

From an implementation perspective, the executable scripts produced by BPDML align with SQL semantics and target low-code platforms such as Bonita. These scripts are automatically validated using unit tests generated with Spock, as outlined in Section~\ref{sec:vdfbpmn:verification:unit_test}. Meanwhile, the formal aspects of process soundness and safety are analyzed using MCMT ver. 3.0, as discussed in Sections~\ref{sec:vdfbpmn:verification:data_aware} and ~\ref{sec:vdfbpmn:verification:guard}.

Furthermore, our approach aims at designing techniques to support business analysts in verifying their process models without requiring expertise in formal modeling techniques or unit testing. The user-friendly validation tools and automated test generation simplify the verification process, allowing business analysts to focus on the business logic and process requirements while ensuring the technical correctness and safety of their models.s

\subsection{Sources of data and their definition}
\label{sec:vdfbpmn:verification:data_source}
As mentioned earlier, the starting point of this framework is BPDML, which supports all the data proposed in data-aware BPMN, except for a single limitation regarding the representation of read-only data. The requirement analysis in Section~\ref{sec:vdfbpmn:related_work} illustrates that BPMN-ProX does not distinguish between read-only and updatable data, treating all data as updatable by default. However, separating read-only data from updatable data is crucial for ensuring data integrity, improving system performance, and reducing the risk of unintended modifications. Therefore, the first step of this work is to extend the data within the activity by introducing a new property to represent read-only data, thereby enabling a more robust classification of data store types.

Each activity, in BDPML~\cite{NourEldin2024LowCode}, has inputs, outputs, and relationships between these inputs and outputs. The inputs and outputs have object types defined as $\mathbf{T}$, where the types of the objects are $\mathbf{T_o} \cup \mathbf{T_d}$, with $\mathbf{T_o}$ is a primitive, system-reserved type (such as strings, integers, boolean, ...) and $\mathbf{T_d}$ is a data type accounting for a semantic domain (like Employee, Payment, Person, ...). The latter refers to the names of database tables used during execution. Additionally, the type of the data (cf. icons attached to the data in Figure~\ref{fig:dfbpmn:dfbpmn:symbols}) is $\mathfrak{T} \in \{data\_object,$ $data\_store,$ $user\_data,$ $system\_data\}$. The definitions of these essential keys in BPDML approach are as follows: activity input, activity output, and finally, the activity itself.

The global inputs are part of the activity data in BPDML, which can be shared among different activities to utilize values allocated within, we extend this input to be more flexible for the separation of the data sources.

\begin{definition}[Global input extension]
\label{def:vdfbpmn:global_input_extension}
A global input extension is defined as $G' = G_{ds} \cup G_{g}$ such that:
\begin{itemize}
    \item $G_{ds}$ is the data store input, which is a tuple $g_{{ds}_i} = (n, \mathbf{to}, A, m, r)$, where:
    \begin{itemize}
        \item $r$ is a boolean variable that equals true if the object is read-only, and false otherwise.
        \item $n$ is the input's name.
        \item $\mathbf{to} \in \mathbf{T}$ is the object type.
        \item $A=\{a_0,\dots,a_{|A|}\}$ is a set of attributes where $a_i=(n', \mathbf{to'})$ such that $n'$ is a name to describe the attribute and $\mathbf{to'} \in \mathbf{T}$ represents the type of the attribute. $g_i.A$ represents to the attributes of $g_i$.
        \item m is a boolean variable that equals $true$ if the object is represent multiple-instance object, and $false$ otherwise.
    \end{itemize}
    \item $G_{g}$ is the global input, which is a tuple $g_{g_i} = (n, \mathbf{to}, \mathfrak{ti}, A, m)$ such that $\mathfrak{ti} \in \mathfrak{T} \setminus \{data\_store\}$ and $n, \mathbf{to}, A, m$ are as defined in $G_{ds}$.
\end{itemize}
\end{definition}

For an activity $act$, $G'(act)$ refers to the set of global inputs, $G'(act)=\{g'_1,\dots, g'_{|G'(act)|}\}$.

\begin{definition}[Activity output extension]
\label{def:vdfbpmn:output_extension}
An activity output is defined as $O' = O_{ds}  \cup O_{g} $ where:
\begin{itemize}
    \item $O_{ds}$ is the data store output, which is a tuple $o_{{ds}_i} = (n, \mathbf{to}, s, A, m, r)$, where $r$ is a boolean variable that equals true if the object is read-only, and false otherwise. The variables $n, \mathfrak{to}, s, A, m$ are as defined in Definition~\ref{def:vdfbpmn:global_input_extension}.
    \item $O_{g}$ is the global output, which is a tuple $o_{g_i} = (n, \mathbf{to}, \mathfrak{ti}, s, A, m)$ such that $\mathfrak{ti} \in \mathfrak{T} \setminus \{data\_store\}$, and $n, \mathbf{to}, s, A, m$ are as defined in Definition~\ref{def:vdfbpmn:global_input_extension}.
\end{itemize}
\end{definition}

For an activity $act$, $O'(act)$ refers to the set of activity outputs extension of $act$, $O'(act)=\{o'_1,\dots,o'_{|O'(act)|}\}$.

\medbreak

\textbf{Persistent data.} 
BPMN-ProX allows to define two types of persistent storages with different access policies. The persistent data are presented by the data of type $\mathfrak{ti} = data\_store$. Formally it defined as follows:

\begin{definition}[Persistent Data]
\label{def:vdfbpmn:persistent_data}
Persistent data is defined as a subset of the inputs and extended outputs of all activities within the process, where the inputs have the type `data\_store'. For a process with a set of activities \( \{act_1, act_2, \dots, act_n\} \):
\[
P = \{ x \mid x \in \bigcup_{i=1}^{n} \left( G_{ds}(act_i) \cup O_{ds}(act_i) \right) \}
\]
with \( G_{ds}(act_i) \) and \( O_{ds}(act_i) \) as defined in Definitions~\ref{def:vdfbpmn:global_input_extension} and~\ref{def:vdfbpmn:output_extension} respectively.

\end{definition}

\begin{figure}[!ht]
    \centering
    \includegraphics[width=0.8\textwidth]{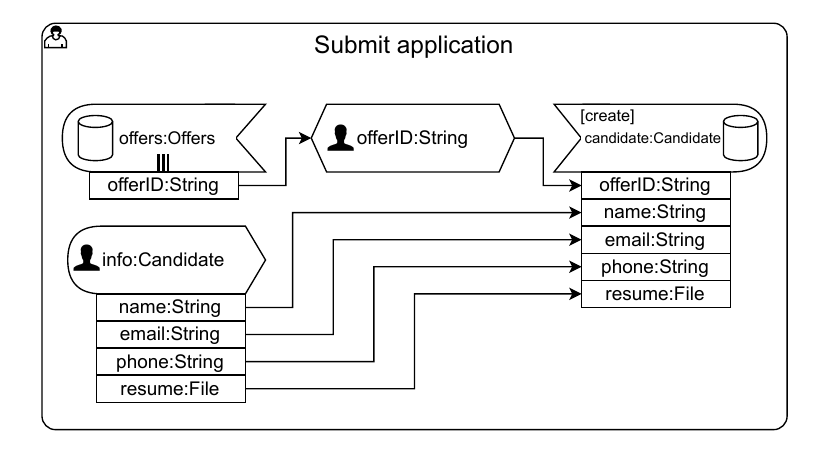}
    \caption{Activity ``Submit application" modeled with BPMN-ProX, showing the interaction between data store input, user input, and internal processing.}
    \label{fig:vdfbpmn:activity_submitapplication}
\end{figure}

According to the definition of persistent data introduced above, $p_1 = (\text{offers}, \text{Offers},$ $\{ (\text{offerID}, \text{String}), $ $ (\text{offerDate}, \text{Date}), (\text{offerStatus}, \text{String}), (\text{description}, \text{String})\}, 1, 1)$ represents a global input used by the process. Since this information originates from a `data\_store' and remains unaltered during process execution, it is part of the persistent data. Consequently, it can be formally included in the set $P$. The user can select from these predefined offers (as shown in Figure~\ref{fig:vdfbpmn:activity_submitapplication}), but cannot modify them—thus, the data store behaves as a read-only input, which aligns with the definition discussed above.

\medbreak

\textbf{Volatile data.} 
BPMN-ProX objects presented as data instance type `data\_object', which are the process variables that used during the execution, and can be filled by user variables or the persistent data. Formally is defined as follows:

\begin{definition}[Volatile Data]
\label{def:vdfbpmn:volatile_data}
Volatile data are defined as a subset of the inputs and extended outputs of all activities within the process, where the inputs have the type `data\_object'. For a process with a set of activities \( \{act_1, act_2, \dots, act_n\} \):
\[
V = \{ x \mid x \in \bigcup_{i=1}^{n} \left( G'(act_i) \cup O'(act_i) \right) \}
\]
with \( G'(act_i) \) and \( O'(act_i) \) as defined in Definitions~\ref{def:vdfbpmn:global_input_extension} and~\ref{def:vdfbpmn:output_extension} respectively.

\end{definition}

\begin{figure}[!ht]
\centering
\includegraphics[width=0.9\textwidth]{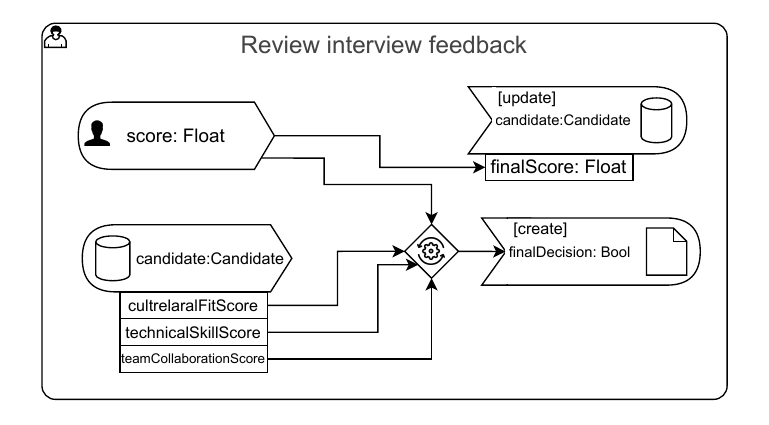}
\centering
\caption{Activity "Review interview feedback" Modeled with BPMN-ProX}
\label{fig:vdfbpmn:activity_reviewinterviewfeedback}
\end{figure}

For instance, $v_1 = (\text{finalDecision}, \text{double}, \emptyset, 0)$ represents a volatile data element used within the activity shown in Figure~\ref{fig:vdfbpmn:activity_reviewinterviewfeedback}. This data is considered volatile because it is created and consumed during the execution of the activity and does not persist beyond it. In this case, `finalDecision' captures the outcome of the interview evaluation (e.g., a score or decision indicator), and it is treated as a temporary, in-memory value. This usage is consistent with the definition of volatile data, which refers to internal variables relevant only to the specific execution context in which they appear. Moreover, such data can be interconnected with other data elements that are eventually stored in a data store.

\subsection{Verification-aware control flow}
The control-flow of a BPMN-ProX process is built by systematically assembling process components that follow the BPMN 2.0 standard. These components are organized using a block-structured approach, meaning that each process is composed of nested blocks with a single entry and a single exit point. This structure guarantees that the flow of the process remains clear, well-defined, and easy to analyze. Typical blocks include patterns like sequences, choices (e.g., exclusive gateways), and parallel executions (e.g., parallel gateways), which can be recursively combined to build complex workflows in a modular and structured way.

We prioritize block-structured BPMN as it facilitates the specification of direct execution semantics for complex constructs such as interrupting exceptions and cancellations, thereby supporting the verification of the resulting models (refer to~\cite{data-aware-BPMN} for more technical details). Despite this focus, our methodology remains applicable even if the control-flow framework is represented using a Petri net, as demonstrated in~\cite{DBLP:conf/bpm/GhilardiGMR20}.

\begin{figure}[!ht]
\centering
\includegraphics[width=1\linewidth]{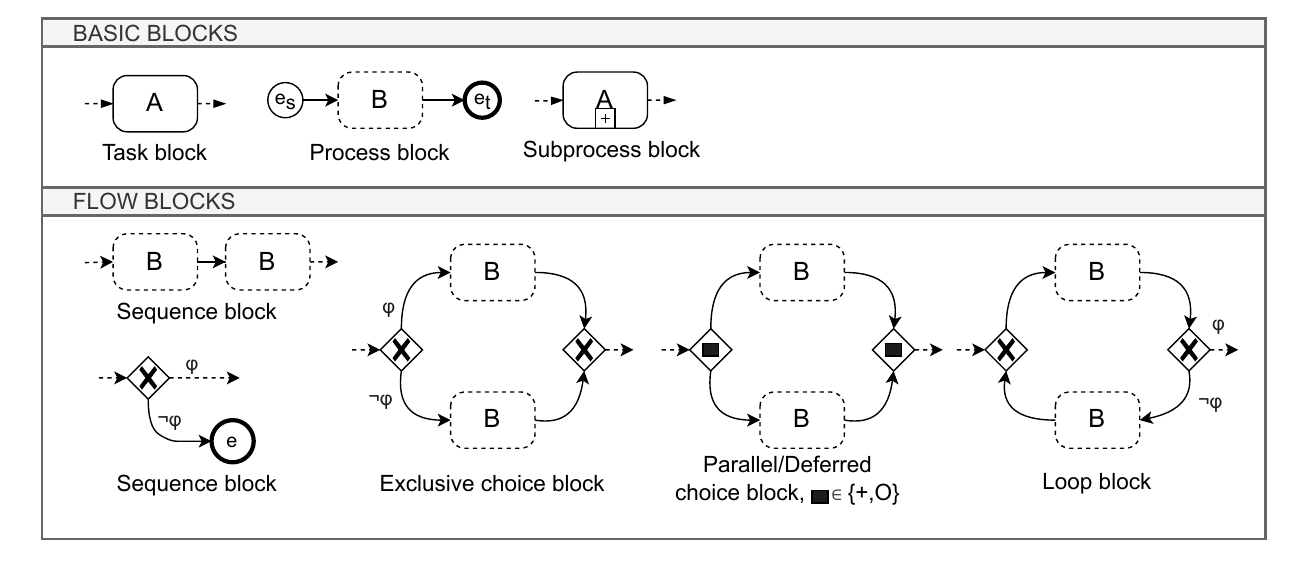}
\caption{Supported BPMN blocks.}
\label{fig:vdfbpmn:supported_block}
\end{figure}

Conceptually, BPMN-ProX supports an equivalent set of blocks as DAB~\cite{data-aware-BPMN}, though its current implementation includes the primary blocks depicted in Figure~\ref{fig:vdfbpmn:supported_block}. Typically, blocks are categorized into leaf blocks (tasks) and non-leaf blocks, which integrate sub-blocks within specific control-flow structures.

Each block implicitly possesses a lifecycle. Initially inactive, a block's state is \textit{idle}. When a process instance reaches an \textit{idle} block during execution, it transitions to the \textit{enabled} state. This transition indicates that the \textit{enabled} block is ready for execution, potentially proceeding non-deterministically based on the discretion of the process executor(s). Once the block is traversed, its lifecycle state changes from \textit{enabled} to \textit{compl}, after which it reverts to the \textit{idle} state, facilitating the progression of the process instance as dictated by the parent block.

The execution rules governing each block type align with BPMN standards. For instance, consider a deferred choice block \textbf{S} with sub-blocks \textbf{B1} and \textbf{B2}. Its lifecycle starts in the \textit{enabled} state, which can advance non-deterministically to the \textit{active} state. This transition enables both \textbf{B1} and \textbf{B2}, changing their states from \textit{idle} to \textit{enabled}. Upon selecting \textbf{B1}, it becomes \textit{active}, while \textbf{B2} reverts to \textit{idle}. Once \textbf{B1} completes and reaches the \textit{compl} state, it transitions back to \textit{idle}, prompting the parent block \textbf{S} to shift from \textit{active} to \textit{compl}.

It is important to note that in the verification phase, we are constrained by certain BPMN components and are required to enforce block-structured models. This restriction ensures that the models remain verifiable and manageable within our framework. Following this logic, the lifecycle model for each block type can be comprehensively defined, maintaining consistency with verification requirements and execution semantics.

\subsection{BPMN-ProX modeling constraints}
\label{sec:vdfbpmn:verification:syntax}
To detect syntax errors before proceeding to the execution phase or transforming BPMN-ProX to MCMT, we have added a syntax error handler to our tool to help users correct these errors. The constraints are defined as follows:

\begin{align}
    &\forall i \in I(act), i.A = \emptyset \implies \exists (i,x) \in DF. \\
    &\forall i \in I.A \implies \exists (i,x) \in DF. \\
    &\forall o \in O, o.s \neq \text{deleted}, i.A = \emptyset \implies \exists (o,x) \in DF.\\
    &\forall o \in O.A  \implies \exists (o,x) \in DF. \\
    &\forall op \in OP \implies \exists (x,op), (op,x') \in DF \text{ and } op.e \neq \emptyset.\\
    &\forall df=(x,y) \in DF \implies x \neq y \text{ and } \mathbf{to}(x) = \mathbf{to}(y) \text{ and } x,y \notin OP.
\end{align}

Where $DF$ is the dataflow relation (see Fig.~\ref{fig:dfbpmn:dfbpmn:symbols}(f)), and $OP$ denotes natural-language-based data processing operators (Fig.~\ref{fig:dfbpmn:dfbpmn:symbols}(h)).

The above constraints ensure syntactic validity across various modeling components. Constraint (6.1) guarantees that every activity input without attributes must have at least one outgoing dataflow connection, preventing isolated tasks. Constraint (6.2) complements this by ensuring that if an activity input has defined attributes, each of these attributes must have an outgoing dataflow connection. Similarly, constraints (6.3) and (6.4) ensure the proper connectivity of dataflow: constraint (6.3) enforces incoming dataflow for data outputs, while constraint (6.4) ensures that output attributes are properly linked. In constraint (6.5), each operation node is required to have at least one incoming and one outgoing connection and must be bound to an executable expression, ensuring that operations are both structurally and semantically valid. Lastly, constraint (6.6) prevents invalid flows by disallowing self-loops and enforcing that the source and destination belong to the same type (e.g., integer, string, etc.), in cases where the dataflow edges do not directly connect operation nodes.

\begin{figure}[!ht]
\centering
\includegraphics[width=0.8\linewidth]{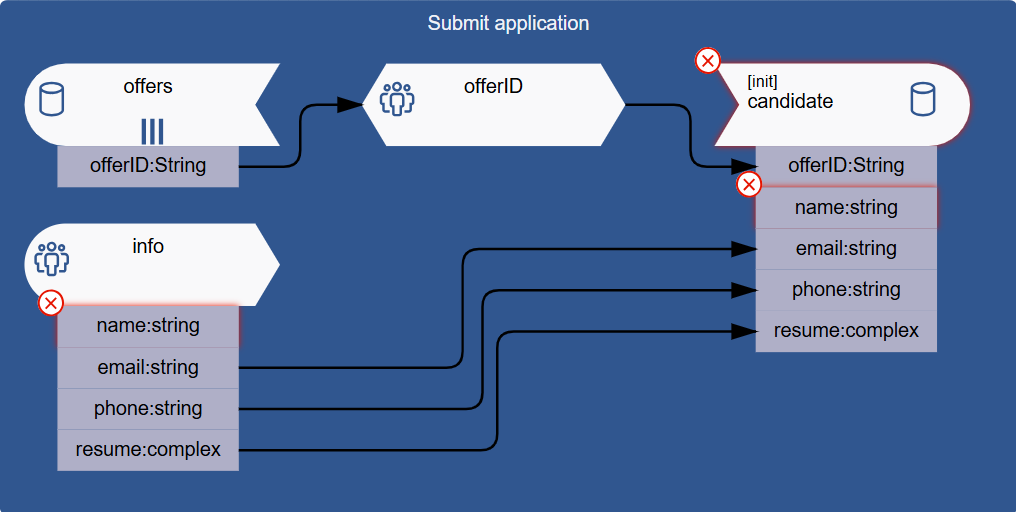}
\caption{Error detection within our tool.}
\label{fig:vdfbpmn:error_handler}
\end{figure}

These constraints are developed within the tool to assist during the modeling phase. For example, Figure~\ref{fig:vdfbpmn:error_handler} illustrates that the attribute (name, string) of Candidate input should be connected to another data instance, and the output attribute (name, string) should be connected to other data.  In particular, name attributes should be interconnected. This helps the user directly detect the missing relation between these two data instances.

\subsection{Encoding BPMN-ProX in data-aware BPMN}
\label{sec:vdfbpmn:verification:data_aware}
The encoding of BPMN-ProX into data-aware BPMN enforces model correctness by translating informal or semi-structured process logic into a formally defined representation where every activity is governed by explicit preconditions and effects. This formalization acts as a constraint mechanism, disallowing the construction of models with undefined data dependencies or inconsistent state transitions. As such, only models that satisfy these semantic rules can be encoded, effectively guaranteeing the correctness of control and data flow. Furthermore, the structure enables formal verification techniques, allowing analysts to detect potential issues—such as data inconsistencies, deadlocks, or unreachable paths—before execution.

We now present how BPMN-ProX can be encoded into data-aware BPMN~\cite{data-aware-BPMN}. We adopt the same definitions of \textit{precondition} and \textit{effect} used in data-aware BPMN. A \textit{precondition} provides possible bindings for the input variables of a task, taking into account persistent data, process data, and external user data. Once a binding for these input variables is selected, the \textit{effect} determines how the task manipulates the process data and updates the persistent data.

A logical approach for querying relational data stored in persistent databases is to use a relational query language such as SQL, which also underpins data-aware BPMN. However, two key considerations must be addressed. First, it is crucial to employ a single query language consistently for different querying requirements, such as defining the preconditions of a task or the conditions for decision-making paths. Second, unlike standard SQL, our queries must accommodate process variables, allowing operations on both persistent and transient data, as well as integration with external data.

Consider a scenario where a candidate submits a job application through an online portal. The system captures the application details and stores them in a process variable. The process then retrieves all relevant information, such as the candidate's qualifications and the job requirements for the selected position. This information is used in the \texttt{WHERE} clause of a \texttt{SELECT} query to fetch specific details from a persistent table that holds data on all applicants and job postings. During the recruitment process, the system may need to query the current state of the candidate’s application or prompt the user (e.g., the hiring manager) to provide feedback after an interview, which is then processed to make a final hiring decision.

\medbreak

\textbf{Preconditions.}
Preconditions specify the circumstances under which a task can be executed. They also retrieve data from the persistent data attached to the leaf block. In BPMN-ProX, the preconditions consist of all inputs and all outputs in the persistent data whose state is set to `read'.

To accommodate these requirements, BPMN-ProX — similarly to PDMML~\cite{DBLP:delta_BPMN} — incorporates a hybrid SQL-based query language that can retrieve persistent data. Consistent with the execution semantics in~\cite{data-aware-BPMN} (which, in turn, align with the customary “variable binding” abstraction in formalisms such as Colored Petri nets), the typical usage of queries in our framework is to return a set of answers from which one is nondeterministically selected to advance the process. The artifact-centric literature has long established this approach as a standard method for handling query results~\cite{DBLP:conf/pods/CalvaneseGM13,DBLP:journals/pvldb/LiDV17,DBLP:journals/mscs/CalvaneseGGMR20}.

\smallskip \noindent \textbf{BPMN-ProX Conditions.}
To define preconditions, we first introduce BPMN-ProX conditions, defined as:
\[
  \mathrm{cond} \;::=\; x_{1} \odot x_{2}
  \quad\big|\quad
  \mathrm{cond}_{1} \;\wedge\; \mathrm{cond}_{2},
\]
where a BPMN-ProX condition is a Boolean expression (with negation pushed inward) over atomic conditions of the form \(x_{1} \odot x_{2}\). Here, \(x_{1}\) and \(x_{2}\) are terms (their shape depends on the context), and
\(\odot \in \{=, \neq, >, <, \le, \ge\}\) is a comparison operator.
We assume component-wise type compatibility of terms (i.e., \(x_{1}\) and \(x_{2}\) must share the same type).
As is customary, the atomic condition \(\texttt{TRUE}\) (which always succeeds) can be introduced as shorthand (similarly for \(\texttt{FALSE}\)).

\smallskip \noindent \textbf{Precondition Queries.}
Using these conditions as building blocks, a BPMN-ProX precondition is defined by the grammar:

\begin{align*}
\text{pre} ::= & \text{query} \\
\text{query} ::= & \text{SELECT } A_1.n, \ldots, A_n.n \text{ FROM } \mathbf{to}(R_1), \ldots, \mathbf{to}(R_m) \text{ WHERE } \text{filter} \\
\text{filter} ::= &  cond \mid \text{TUPLE}(\vec{x}) \text{ IN } R.A.n \mid \text{filter}_1 \textbf{ AND } \text{filter}_2 \\
\end{align*}

Each $\mathbf{to}(R_i)$ in the SQL-like query represents the read persistent data relation. The terms used in the \text{cond} of \text{filter} match the process variables, constants, or user data declared within the same leaf block. Additionally, they can utilize attributes specified in the \text{FROM} clause of the query expression (i.e., $A_1, \ldots, A_n$). When formulating queries, the notation $R.A$ can be employed to explicitly denote attribute $A$ of table $P$, where $\{r_0,\dots, r_k\} \in R \subseteq P$ such that $ r_i.s = 'read' \mid r_i \in I$.

In the recruitment process scenario touched in Figure~\ref{fig:vdfbpmn:activity_submitapplication} represent an activity of BPMN-ProX, the following query can be used to list offer id of the offers that exists in the database:
\begin{align*}
\texttt{SELECT offerID FROM Offers}
\end{align*}

\medbreak

\textbf{Effects.}
Task effects consist of BPMN-ProX data manipulation statements that operate over process variables and updatable persistent data. In this context, an \textit{input variable} refers to user data or attributes from the precondition attached to the same leaf block as the effect. Any process variable \texttt{v} can be updated using a simple assignment statement \texttt{v = u}, where \texttt{u} $\in V$ such that \texttt{u} $\in O$, and \texttt{u} is either a constant or an input variable. Graphically, this corresponds to all outputs whose type is \texttt{data\_object}. We assume that each process variable can be assigned at most once in a single update. For instance, in Figure~\ref{fig:vdfbpmn:activity_reviewinterviewfeedback} `finalScore' attribute should be updated to the value of `score' provided by the user.

We also allow insertion and deletion of tuples in updatable persistent storage:

\medbreak
\smallskip \noindent \textbf{Insertion.}

An \textit{insertion} (statement) on some updateable persistent relation \texttt{U} is defined as \texttt{INSERT v\_1, \ldots, v\_n INTO U.n}, where each \texttt{v\_i} is either a constant, a process variable, or an input variable. $U \in P$ where $U.s = \text{'init'  and } U.r = false$.
An automatic transformation of the graphical notation to this definition to be compatible with SQL query. This \texttt{INSERT} statement is similar to the corresponding classical DML (data manipulation language) statement in SQL. However, it deliberately avoids using the \texttt{VALUES} clause since we insert one tuple at a time, and so we can rely on the more compact notation where the elements to be inserted are directly indicated close to \texttt{U}. For example, in Figure~\ref{fig:vdfbpmn:activity_submitapplication}, the `candidate' data output is in `create' state, that mean that we need to insert into `Candidate' table based on the user data.

\smallskip \noindent \textbf{Deletion.}

A \textit{deletion} statement is handled analogously: the only difference is that the output’s state is set to `deleted'.

\smallskip \noindent \textbf{Conditional Updates.}

We also allow to perform conditional updates. In BPMN-ProX, the conditional update was defined using gherkin syntax which is converted automatically to SQL syntax. The gerkin syntax already supported in BDPML to express the behavoir of the data operation. For this transformation, a modified SQL \texttt{CASE} statement is employed directly embedded into the update logic. This statement logically resembles an \texttt{if-then-else} expression with multiple \texttt{else-if} branches, and in which each condition in the \texttt{if} part is a query. To ensure verifiability \cite{data-aware-BPMN,DBLP:delta_BPMN,DBLP:conf/pods/CalvaneseGM13}, it is necessary for the statement to obey to one limitation: it cannot access any other table beyond the one that is being updated. The conditional update statement has the format:

\begin{align*}
    \text{\textbf{UPDATE} U.n \textbf{SET} } R.A_1.n=@v_1, ..., R.A_m.n=@v_m  \textbf{ WHERE } \\
    \textbf{CASE WHEN} F_1  \textbf{ THEN }  @v_1=u_1^1, ..., @v_m=u_m^1 \\ 
    \textbf{WHEN } F_k \textbf{ THEN}  @v1=u_k^1, ..., @v_m=u_m^k\\
     ...\\
    \textbf{ELSE } @v_1=u_1^c, ..., @v_m=u_m^c
\end{align*}

This statement is the most sophisticated one in the offered language as it requires the modeler to take care of the following two aspects. First, similarly to the SQL's \texttt{UPDATE} statement, which can modify multiple tuples in a table, ours performs a (conditional) \textit{bulk edit} of elements in each tuple of \texttt{U}, and the \texttt{SET} clause specifies (using names of the attributes of \texttt{U} with the \texttt{U}'s name in the prefix) what are exactly those elements. The \texttt{SET} clause also uses placeholder variables \texttt{@vi} that support the conditional update logic: whenever a tuple in \texttt{U} satisfies one of the \texttt{Fi} filters, the corresponding \texttt{THEN} clause will assign concrete values \texttt{$u^j_i$} to all the placeholder variables mentioned in \texttt{SET}. Second, the modeler has to carefully control the variables and attributes used both in the \texttt{WHEN} and \texttt{THEN} clauses. As already mentioned, each \texttt{Fi} cannot access updatable persistent relations but \texttt{U} itself. At the same time, it can reuse elements from the precondition query such as variables and attributes. This, in turn, allows to use \texttt{Fi} for filtering results returned by the precondition query, and thus allowing to carefully select the data that are going to be used in the final update of every single tuple of \texttt{U}. As for the elements appearing in \texttt{THEN} clauses, their values can be constants as well as elements taken from results returned by the precondition query. 

\medbreak

The overall execution semantics of leaf blocks is defined as follows. Once a leaf block is \textit{enabled}, the modeler can provide bindings for its newly introduced attributes. If, under these bindings, the precondition holds for at least one binding of its attributes, then the leaf block may (nondeterministically) fire, depending on the choice of the process executors. Upon firing, the bindings of precondition attributes and newly introduced variables provide a grounding for the effect. Once the effect is applied, the block completes its execution and the leaf block’s lifecycle state becomes \textit{compl}, as described earlier.

A final requirement is that if a task has both a precondition and an effect, it is considered atomic in terms of data updates. While not strictly a technical limitation, this requirement ensures that insertions, deletions, or updates are applied to the same data snapshot used when the precondition was verified, adhering to standard \textit{transactional semantics}. Violating atomicity could lead to race conditions with other specifications updating the same process variables or persistent tables. Note that race conditions can still occur at the process level, for example when parallel branches are active. Consequently, enforcing atomicity on leaf blocks does not reduce the overall generality of our approach.

\subsection{Guards for conditional flows}
\label{sec:vdfbpmn:verification:guard}
Building on the concept of task effects and preconditions, BPMN-ProX also facilitates conditional decision-making within process flows through the use of guards, which govern the routing of process instances based on specific conditions.

BPMN-ProX statements are essential in scenarios where blocks employ choice splits to conditionally route process instances. In these cases, each conditional flow is associated with a BPMN-ProX condition that consists of process variables or constants. Notably, restricting conditions to process variables does not pose a limitation, as these variables can be populated with data from data stores or external environments. These conditions are expressed as textual statements without graphical notation since BPMN-ProX modeling emphasizes activities rather than gateways.

As illustrated in Figure~\ref{fig:vdfbpmn:supported_block}, each choice split is assumed to provide two outputs with complementary guards. Consequently, the user needs to specify only one guard, \( \phi \), while the complementary guard, denoted as \( \neg \phi \) in the figure, is automatically generated through the syntactic manipulation of \( \phi \). This process involves applying De Morgan's laws until the negation appears directly before atomic conditions, followed by substituting the negated atomic conditions with their complementary counterparts (e.g., replacing \( \leq \) with \( > \)).

\subsection{Automating unit tests}
\label{sec:vdfbpmn:verification:unit_test}
As introduced in Section~\ref{sec:vdfbpmn:verification:data_aware}, the transformation of BPMN-ProX into data-aware BPMN is designed to enable formal verification by converting BPMN-ProX models into a format compatible with the MCMT model checker. However, this transformation currently lacks support for complex operations (such as arithmetic operations, conditions, and loops), users must manually validate these scripts. To address this challenge, we propose automating unit tests for the generated code. This automation assists users in adding and testing specific cases that are not covered by the automatically generated tests.

\begin{figure}[!ht]
\centering
\includegraphics[width=1\linewidth]{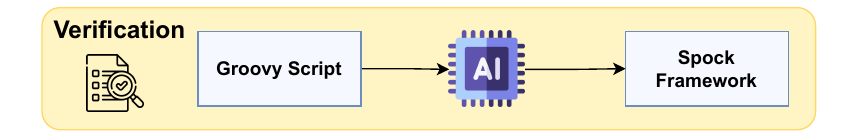}
\caption{Automated unit testing pipeline.}
\label{fig:vdfbpmn:poc:unittest}
\end{figure}

Figure~\ref{fig:vdfbpmn:poc:unittest} illustrates the automated unit testing pipeline. The process begins with the Groovy Script, which contains the logic generated from the BPMN-ProX transformation. This script is then passed to an AI-powered large language model (LLM) that leverages prompt engineering~\cite{prompt-engineering} to analyze the code. The LLM dynamically generates unit tests designed to cover a wide range of input scenarios, including both typical cases and edge cases. Once generated, these unit tests are integrated into the Spock Framework, a testing and specification framework for Groovy applications. The Spock Framework executes the tests, providing detailed verification of the script's correctness. This pipeline ensures a streamlined, automated approach to validating complex operations, significantly reducing the need for manual intervention.

For example, consider an activity that evaluates a candidate’s suitability for a role based on multiple criteria. The data operation first checks if the candidate has a technical skills score. If they do, it’s generally preferred that the score is above 3.5. After this, the evaluation considers additional aspects such as cultural fit, teamwork, and overall performance. Ideally, the total score should be at least 11, but the broader context is also taken into account. If all factors align well, the candidate may proceed to the next stage of the hiring process; otherwise, they might not be the right fit for the role. An example of a generated unit test verifying this evaluation logic is presented in the Listing~\ref{lst:vdfbpmn:unit-test}.

\begin{lstlisting}[style=groovytest, caption={Generated unit test verifying candidate selection logic}, breaklines=true, breakatwhitespace=true, label={lst:vdfbpmn:unit-test}]
import spock.lang.Specification
import spock.lang.Unroll

class CandidateSelectionSpec extends Specification {
    def "should determine if candidate proceeds to the next stage based on technical score and combined assessment"() {
        expect:
        def result = isCandidateSelected(technicalScore, culturalFit, teamCollaboration, finalEvaluation)
        result == expectedOutcome
        where:
        technicalScore | culturalFit | teamCollaboration | finalEvaluation || expectedOutcome
        4.0            | 4.0         | 4.0               | 4.0             || true
        3.6            | 3.5         | 4.0               | 3.5             || true
        3.5            | 5.0         | 4.0               | 4.0             || false
        4.5            | 3.0         | 3.0               | 3.0             || false
        2.5            | 5.0         | 5.0               | 5.0             || false
        3.9            | 3.0         | 3.5               | 4.0             || true
        4.2            | 2.0         | 2.5               | 3.0             || false
        3.8            | 4.0         | 4.0               | 3.0             || true
        4.0            | 2.0         | 2.0               | 6.5             || true
        3.4            | 4.0         | 4.0               | 4.0             || false
    }

    boolean isCandidateSelected(double technicalScore, double culturalFit, double teamCollaboration, double finalEvaluation) {
        return technicalScore > 3.5 && (culturalFit + teamCollaboration + finalEvaluation) >= 11
    }
}

\end{lstlisting}

The code is generated through the transformation of natural language/Gherkin into a Groovy script, as discussed in \cite{NourEldin2024LowCode}. Lines 1–10 correspond to the Spock framework structure and are automatically generated by the LLM. Lines 11–20 contain the generated test cases, which the citizen developer can easily extend to verify whether the transformation from Gherkin to code aligns with user requirements. The user can easily run the code and check whether any errors exist\footnote{An open-source web console for running Groovy scripts: \url{https://github.com/groovy-console/groovy-web-console}}.

\section{Proof of Concept}
\label{sec:verification_techniques}
\label{sec:vdfbpmn:poc}
To ensure the verifiability of BPMN-ProX processes within our tool, we developed a translator that processes the chosen model according to the previously established modeling guidelines~\footnote{The implementation is available on the tool website: \url{https://github.com/NourEldin-Ali/BPMN-ProX}.}. This translator converts models into a syntax compatible with a cutting-edge model checker, capable of parametrically verifying data-aware processes with read-only relations, specifically utilizing the latest version of MCMT \cite{data-aware-BPMN,DBLP:conf/birthday/CalvaneseGGMR19,DBLP:journals/mscs/CalvaneseGGMR20}.

The translation process begins by confirming if the input model is block-structured and isolating the various blocks using a traversal algorithm of independent interest. Each block is then individually transformed into corresponding MCMT instructions, implementing the encoding mechanism detailed in \cite{data-aware-BPMN} rule by rule. This method is effective because the BPMN-ProX syntax for data definition and manipulation closely mirrors the abstract, logical language used in the referenced works.

For verification, we specify the properties to be checked, defining each property as a condition identifying an undesirable state of the model. To add a property, we use our BPMN-ProX extension elements, incorporating a reserved identifier \texttt{verify} to add property key-value pairs directly to the process. For instance, to verify the safety of the model in Fig.~\ref{fig:dfbpmn:dfbpmn_full_example}, one might write the BPMN-ProX condition (status=Archived AND interviewOutcome=Rejected) to ensure the hiring application process concludes with the candidate being rejected (as indicated by the End event "Candidate Rejected" in Fig.~\ref{fig:dfbpmn:dfbpmn_full_example}), resulting in the application being archived. A special variable \texttt{lifecycleHiring} is used to access the process lifecycle state. Generally, querying the process lifecycle involves using an internal variable \texttt{lifecycleModelName}, where \texttt{ModelName} is the actual process model name. Verification of lifecycle properties for individual blocks can be managed by introducing dedicated case variables and manipulating them in accordance with the block's lifecycle evolution.

It is noteworthy that, although not explicitly reflected in the BPMN-ProX language, BPMN-ProX supports modeling and verification of multi-instance scenarios where process instances can access and manipulate the same data store. Formal details of data-aware BPMN are provided in \cite{data-aware-BPMN}. Summarizing \cite{data-aware-BPMN}, it is indicated that an unlimited number of simultaneously active process instances can be verified for safety if they do not explicitly reference each other. Explicit mutual references can be managed if the maximum number of active process instances is known a priori.

\bigbreak

\begin{figure}[!ht]
\centering
\includegraphics[width=1\linewidth]{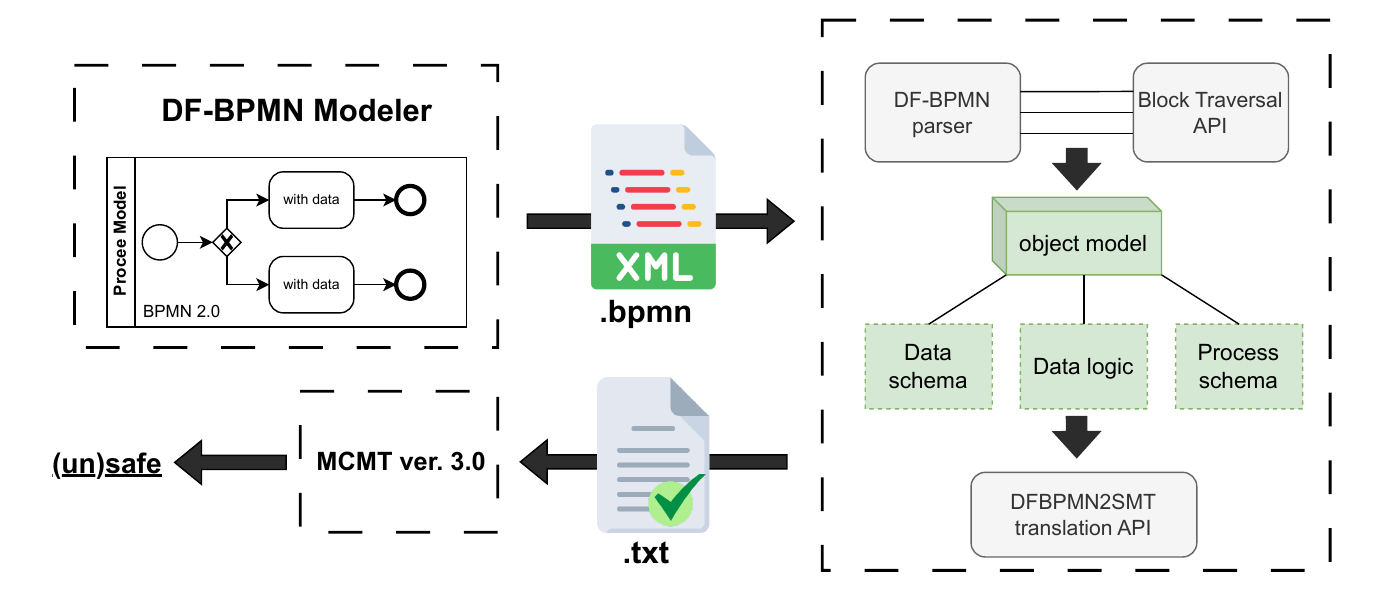}
\caption{Conceptual architecture of the BPMN-ProX framework.}
\label{fig:vdfbpmn:poc}
\end{figure}

Figure~\ref{fig:vdfbpmn:poc} illustrates the overall verification toolchain. Initially, a modeler creates a BPMN-ProX process by enhancing a regular block-structured BPMN 2.0 process with a BPMN-ProX specification. The modeler then exports the BPMN-ProX process as an XML-formatted .bpmn file. This file is processed by our Java-based tool, which utilizes APIs to generate a process specification verifiable by MCMT\footnote{\url{http://users.mat.unimi.it/users/ghilardi/mcmt/}}. Essentially, the tool performs two major steps to process the BPMN-ProX model. Firstly, it employs our BPMN-ProX extension API to access process components from the .bpmn file, using our block traversal API and BPMN-ProX parser to identify blocks and BPMN-ProX statements/declarations, generating BPMN-ProX objects accordingly. These objects are specified based on an object model derived from the formalism in \cite{data-aware-BPMN}, encompassing a data schema for case variables and relations (R and C), a process schema for nested process block definitions, and a data logic for update declarations and conditions assigned to blocks. The block traversal API implements an algorithm to detect nested blocks conforming to the object model structure. Through the DFBPMNX2SMT translation API, which follows the translation method in \cite{data-aware-BPMN}, the tool processes the object model and generates a text file with the BPMN-ProX process specification in MCMT syntax.

Finally, the derived specification can be checked directly using the MCMT tool, which determines whether the specification is safe or unsafe concerning the specified "bad" property. MCMT can be executed via command line with: mcmt  [--time] . The [--time] argument is optional and used to display the MCMT execution time. Additional details on model checker installation, safety property specification, advanced execution options.


\section{Conclution}
\label{sec:conclusion}
In this paper, BPMN-ProX is introduced, which achieves an interesting trade-off between expressiveness and the possibility of applying sophisticated parameterized verification techniques to ascertain the safety of the produced models. This approach merges the concepts of low-code modeling and theoretical process modeling verification, allowing users to model, validate, verify, and execute a BPMN without technical experimentation. Future work aims to make the modeling language more intuitive by integrating artificial intelligence techniques, enabling users to model and execute a BPMN more effectively. Additionally, future research should focus on enhancing dynamic theoretical verification methods, particularly regarding database-centric verification and process-level verification, to comprehensively address the complexities and nuances of business process validation. Indeed, we will also focus on automating integrated tests for verifying executable activity paths based on data, further ensuring robustness and correctness.

\bibliographystyle{splncs04}
\bibliography{biblio}

\begin{thebibliography}{10}
\providecommand{\url}[1]{\texttt{#1}}
\providecommand{\urlprefix}{URL }
\providecommand{\doi}[1]{https://doi.org/#1}

\bibitem{DBLP:journals/corr/BohmerR15}
B{\"{o}}hmer, K., Rinderle{-}Ma, S.: A systematic literature review on process model testing: Approaches, challenges, and research directions. CoRR  \textbf{abs/1509.04076} (2015)

\bibitem{data-aware-BPMN}
Calvanese, D., Ghilardi, S., Gianola, A., Montali, M., Rivkin, A.: Formal modeling and smt-based parameterized verification of data-aware {BPMN}. In: Business Process Management - 17th International Conference, {BPM} 2019, Vienna, Austria, September 1-6, 2019, Proceedings. Lecture Notes in Computer Science, vol. 11675, pp. 157--175. Springer (2019)

\bibitem{DBLP:conf/birthday/CalvaneseGGMR19}
Calvanese, D., Ghilardi, S., Gianola, A., Montali, M., Rivkin, A.: From model completeness to verification of data aware processes. In: Description Logic, Theory Combination, and All That - Essays Dedicated to Franz Baader on the Occasion of His 60th Birthday. Lecture Notes in Computer Science, vol. 11560, pp. 212--239. Springer (2019)

\bibitem{DBLP:journals/mscs/CalvaneseGGMR20}
Calvanese, D., Ghilardi, S., Gianola, A., Montali, M., Rivkin, A.: Smt-based verification of data-aware processes: a model-theoretic approach. Math. Struct. Comput. Sci.  \textbf{30}(3),  271--313 (2020)

\bibitem{DBLP:conf/pods/CalvaneseGM13}
Calvanese, D., Giacomo, G.D., Montali, M.: Foundations of data-aware process analysis: a database theory perspective. In: Proceedings of the 32nd {ACM} {SIGMOD-SIGACT-SIGART} Symposium on Principles of Database Systems, {PODS} 2013, New York, NY, {USA} - June 22 - 27, 2013. pp. 1--12. {ACM} (2013)

\bibitem{DBLP:conf/caise/CalvaneseMPR19}
Calvanese, D., Montali, M., Patrizi, F., Rivkin, A.: Modeling and in-database management of relational, data-aware processes. In: Advanced Information Systems Engineering - 31st International Conference, CAiSE 2019, Rome, Italy, June 3-7, 2019, Proceedings. vol. 11483, pp. 328--345. Springer (2019)

\bibitem{bpmn_specification}
Chinosi, M., Trombetta, A.: {BPMN:} an introduction to the standard  (2012)

\bibitem{DBLP:activity-view}
Combi, C., Oliboni, B., Weske, M., Zerbato, F.: Conceptual modeling of processes and data: Connecting different perspectives. In: Conceptual Modeling - 37th International Conference, {ER} (2018)

\bibitem{DBLP:journals/siglog/DeutschHLV18}
Deutsch, A., Hull, R., Li, Y., Vianu, V.: Automatic verification of database-centric systems. {ACM} {SIGLOG} News  \textbf{5}(2),  37--56 (2018)

\bibitem{DBLP:conf/wrla/DuranRS18}
Dur{\'{a}}n, F., Rocha, C., Sala{\"{u}}n, G.: Symbolic specification and verification of data-aware {BPMN} processes using rewriting modulo {SMT}. In: Rewriting Logic and Its Applications - 12th International Workshop, {WRLA} 2018, Held as a Satellite Event of ETAPS, Thessaloniki, Greece, June 14-15, 2018, Proceedings. Lecture Notes in Computer Science, vol. 11152, pp. 76--97. Springer (2018)

\bibitem{DBLP:BAUML}
Esta{\~{n}}ol, M., Sancho, M., Teniente, E.: Ensuring the semantic correctness of a {BAUML} artifact-centric {BPM}. Inf. Softw. Technol.  \textbf{93},  147--162 (2018)

\bibitem{DBLP:conf/bpm/GhilardiGMR20}
Ghilardi, S., Gianola, A., Montali, M., Rivkin, A.: Petri nets with parameterised data - modelling and verification. In: Business Process Management - 18th International Conference, {BPM} 2020, Seville, Spain, September 13-18, 2020, Proceedings. Lecture Notes in Computer Science, vol. 12168, pp. 55--74. Springer (2020)

\bibitem{DBLP:delta_BPMN}
Ghilardi, S., Gianola, A., Montali, M., Rivkin, A.: Delta-bpmn: {A} concrete language and verifier for data-aware {BPMN}. In: Business Process Management - 19th International Conference, {BPM} 2021 (2021)

\bibitem{DBLP:conf/caise/GiacomoOET17}
Giacomo, G.D., Oriol, X., Esta{\~{n}}ol, M., Teniente, E.: Linking data and {BPMN} processes to achieve executable models. In: Advanced Information Systems Engineering - 29th International Conference, CAiSE 2017, Essen, Germany, June 12-16, 2017, Proceedings. Lecture Notes in Computer Science, vol. 10253, pp. 612--628. Springer (2017)

\bibitem{DBLP:journals/bpmj/Guerreiro21}
Guerreiro, S.: Conceptualizing on dynamically stable business processes operation: a literature review on existing concepts. Bus. Process. Manag. J.  \textbf{27}(1),  24--54 (2021)

\bibitem{DBLP:conf/otm/Hull08}
Hull, R.: Artifact-centric business process models: Brief survey of research results and challenges. In: On the Move to Meaningful Internet Systems: {OTM} 2008, {OTM} 2008 Confederated International Conferences, CoopIS, DOA, GADA, IS, and {ODBASE} 2008, Monterrey, Mexico, November 9-14, 2008, Proceedings, Part {II}. Lecture Notes in Computer Science, vol.~5332, pp. 1152--1163. Springer (2008)

\bibitem{DBLP:journals/pvldb/LiDV17}
Li, Y., Deutsch, A., Vianu, V.: {VERIFAS:} {A} practical verifier for artifact systems. Proc. {VLDB} Endow.  \textbf{11}(3),  283--296 (2017)

\bibitem{DBLP:conf/bpm/MeyerPFW13}
Meyer, A., Pufahl, L., Fahland, D., Weske, M.: Modeling and enacting complex data dependencies in business processes. In: Business Process Management - 11th International Conference, {BPM} 2013, Beijing, China, August 26-30, 2013. Proceedings. Lecture Notes in Computer Science, vol.~8094, pp. 171--186. Springer (2013)

\bibitem{DBLP:dbnets}
Montali, M., Rivkin, A.: Db-nets: On the marriage of colored petri nets and relational databases. Trans. Petri Nets Other Model. Concurr.  \textbf{12},  91--118 (2017)

\bibitem{NourEldin2024LowCode}
Nour~Eldin, A., Baudot, J., Dalmas, B., Gaaloul, W.: Low-code solutions for business process dataflows: From modeling to execution (2024), manuscript under review. Available at SSRN

\bibitem{dfbpmn}
{Nour Eldin}, A., Baudot, J., Gaaloul, W.: Zooming in for clarity: Towards low-code modeling for activity data flow. In: Business Process Management Forum - {BPM} 2023 Forum, Utrecht, The Netherlands, September 11-15, 2023, Proceedings. vol.~490, pp. 267--282. Springer (2023)

\bibitem{omg2011umls}
OMG: {OMG Unified Modeling Language (OMG UML), Superstructure, Version 2.4.1} (August 2011), \url{http://www.omg.org/spec/UML/2.4.1}

\bibitem{DBLP:conf/ant/PaivaFFM18}
Paiva, A.C.R., Flores, N.H., Faria, J.P., Marques, J.M.G.: End-to-end automatic business process validation. In: The 9th International Conference on Ambient Systems, Networks and Technologies {(ANT} 2018) / The 8th International Conference on Sustainable Energy Information Technology {(SEIT} 2018) / Affiliated Workshops, May 8-11, 2018, Porto, Portugal. Procedia Computer Science, vol.~130, pp. 999--1004. Elsevier (2018)

\bibitem{DBLP:conf/caise/PolyvyanyyWOB19}
Polyvyanyy, A., van~der Werf, J.M.E.M., Overbeek, S., Brouwers, R.: Information systems modeling: Language, verification, and tool support. In: Advanced Information Systems Engineering - 31st International Conference, CAiSE 2019, Rome, Italy, June 3-7, 2019, Proceedings. Lecture Notes in Computer Science, vol. 11483, pp. 194--212. Springer (2019)

\bibitem{DBLP:conf/otm/Reichert12}
Reichert, M.: Process and data: Two sides of the same coin? In: On the Move to Meaningful Internet Systems: {OTM} 2012, Confederated International Conferences: CoopIS, DOA-SVI, and {ODBASE} 2012, Rome, Italy, September 10-14, 2012. Proceedings, Part {I}. Lecture Notes in Computer Science, vol.~7565, pp. 2--19. Springer (2012)

\bibitem{prompt-engineering}
Sahoo, P., Singh, A.K., Saha, S., Jain, V., Mondal, S., Chadha, A.: A systematic survey of prompt engineering in large language models: Techniques and applications. CoRR  \textbf{abs/2402.07927} (2024)

\bibitem{DBLP:SLR}
Steinau, S., Marrella, A., Andrews, K., Leotta, F., Mecella, M., Reichert, M.: {DALEC:} a framework for the systematic evaluation of data-centric approaches to process management software. Softw. Syst. Model.  \textbf{18}(4),  2679--2716 (2019)

\end{thebibliography}

\end{document}